\documentclass[12pt,preprint]{aastex}

\def\lea{\mathrel{<\kern-1.0em\lower0.9ex\hbox{$\sim$}}}
\def\gea{\mathrel{>\kern-1.0em\lower0.9ex\hbox{$\sim$}}}
\newcommand{\lta}{{\>\rlap{\raise2pt\hbox{$<$}}\lower3pt\hbox{$\sim$}\>}}
\newcommand{\gta}{{\>\rlap{\raise2pt\hbox{$>$}}\lower3pt\hbox{$\sim$}\>}}

\begin{document}

\title{New Constraints on Mass-Dependent Disruption of Star Clusters in M51}
\author{Rupali Chandar,\altaffilmark{1}
Bradley C. Whitmore,\altaffilmark{2}
Daniela~Calzetti,\altaffilmark{3}
Daiana Di~Nino,\altaffilmark{4}\\
Robert C. Kennicutt,\altaffilmark{5}
Michael Regan,\altaffilmark{2} and
Eva Schinnerer\/\altaffilmark{6}
}

\shortauthors{Chandar et al.}
\email{Rupali.Chandar@utoledo.edu}
\altaffiltext{1}{Department of Physics \& Astronomy, The University of Toledo, Toledo, OH 43606, USA}
\altaffiltext{2}{Space Telescope Science Institute, Baltimore, MD 21218, USA}
\altaffiltext{3}{Dept.\ of Astronomy, University of Massachusetts, Amherst, MA 01003, USA}
\altaffiltext{4}{CICLOPS, Space Science Institute, 4750 Walnut Street, Boulder, CO 80301, USA}
\altaffiltext{5}{Institute of Astronomy, Cambridge University, Cambridge, UK}
\altaffiltext{6}{Max-Planck-Institut f\"{u}r Astronomie, K\"{o}nigstuhl 17, D-69117 Heidelberg, Germany}

\begin{abstract}

We use \textit{UBVI}H$\alpha$ images of the Whirlpool galaxy, M51, taken with the ACS and WFPC2 cameras  on the {\em Hubble Space Telescope} 
(\textit{HST}) to select star clusters, and to estimate their masses and ages by
comparing their observed colors with predictions from population synthesis models. We construct the mass function of intermediate age (1--$4\times 
10^8$~yr) clusters, and find that it is well described by a power law, $\psi(M) \propto M^{\beta}$, with $\beta=-2.1\pm0.2$, for clusters more massive
than $M\approx6\times10^3~M_{\odot}$. This extends the mass function of intermediate age clusters in M51 to masses lower by nearly a factor of five over previous determinations. The mass function does not show evidence for curvature at either the high or low mass end. This shape indicates that there is no evidence for the earlier disruption of lower mass clusters compared with their higher mass counterparts (i.e., no mass-dependent disruption) over the observed range of masses and ages, or for a physical upper mass limit $M_C$ with which clusters in M51 can form. These conclusions differ from previous suggestions based on poorer-quality \textit{HST} observations.
We discuss their implications for the formation and disruption of the clusters.
Ages of clusters in two ``feathers,'' stellar features extending from the outer portion of a spiral arm, show that the feather with a larger pitch angle formed earlier, and over a longer period, than the other.

\end{abstract}


\section{INTRODUCTION}

The disruption of star clusters plays an important role in building up the stellar populations of galaxies. While most stars form in groups and clusters, the majority of these clusters disrupt quickly, releasing their remaining stars to the field. However the details of how clusters disrupt, e.g., the physical processes that dominate disruption on different timescales are still controversial. We have previously suggested that the early disruption of clusters for the first $\tau \lea \mbox{few}\times10^8$~yr, does not
depend strongly on the mass of the clusters or on the type of host galaxy.
In this ``quasi-universal'' picture, the distribution of cluster masses and ages $g(M,\tau)$ is quite similar from galaxy to galaxy, although it may be modulated somewhat by different densities of molecular material within galaxies for ages $\tau \gea 10^8$~yr see e.g., Fall et~al.\ 2005, 2009; Whitmore et~al.\ 2007; Chandar et~al.\ 2010a for details). Variations in the rate of cluster formation will also affect $g(M,\tau)$, even if the disruption rate is identical in different galaxies, although the observational evidence
so far suggests that these variations are modest (see discussion in Chandar et~al.\ 2010b). A different picture, where the time it takes a cluster to disrupt
depends on both its mass and on its host galaxy, has been suggested by Lamers et~al.\ (2005); we will refer to this as the ``galaxy-dependent'' picture.

Observations of the cluster system in M51 have played a central role in the
development of the  galaxy-dependent picture of cluster disruption. Assuming that clusters disrupt as a power-law with mass, such that the disruption timescale $\tau_d=\tau_* (M/10^4)^k$, where $\tau_*$ is the characteristic time it takes a $10^4~M_{\odot}$ cluster to disrupt, Boutloukos \& Lamers (2003), Lamers et~al.\ (2005), and Gieles et~al.\ (2005) suggested that clusters in M51 disrupt with parameters $k\approx0.6$ and $\tau_*\approx1$--$2\times10^8$~yr, i.e., that clusters in M51 disrupt on fairly short timescales. These results were based on masses and ages of clusters determined from two pointings with WFPC2 on \textit{HST} (see e.g,
Figure~1 in Bastian et~al.\ 2005).

By contrast, it was suggested that clusters in the Magellanic Clouds have significantly longer disruption times than in M51, with $\tau_*\approx8\times10^9$~yr (Boutloukos \& Lamers 2003; Lamers et~al.\ 2005; de~Grijs \& Anders 2006). In the galaxy-dependent picture, these strong variations in $\tau_*$ are explained by differences in the strength of the tidal field (Lamers et~al.\ 2005) and in the density of giant molecular clouds (Gieles 
et~al.\ 2006a) from galaxy to galaxy.

Recently, we did not find evidence for mass-dependent disruption of clusters in the Magellanic Clouds, and we found that the value of $\tau_*\approx8\times10^9$~yr suggested previously implies observed masses and ages which are below those available in current cluster catalogs (e.g., see Figures~3 and 5 in Chandar et~al.\ 2010a). Similarly, we did not find observational evidence for the mass-dependent disruption of clusters (with ages $\tau \lea 4\times10^8$~yr) in the nearby spiral galaxy M83 (Chandar et~al.\ 2010b), despite the fact that it has a similar morphology, metallicity, and molecular 
mass as M51.\footnote{M83 and M51 have a similar metallicity (see Bresolin et~al.\ 2002 for M83 and Moustakas et~al.\ 2010 for M51). The total mass of molecular gas in M51 ($2\times10^9M_{\odot}$; Schuster et~al.\ 2007)
is within a factor of two of the mass of molecular gas in M83 ($3.9\times10^9~M_{\odot}$; Lundgren et~al.\ 2003).}

Given the suggestion of a short disruption time for clusters in M51, and the critical role this plays in the galaxy-dependent disruption picture, it is important to check the previous results. Here, we determine masses and ages
of star clusters in M51 using the best available \textit{HST} observations,
which are deeper, have better spatial coverage, and higher angular resolution
than used in the works mentioned above, in order to assess the validity of the
previous claims. We focus on constructing the mass function of intermediate-age (1--$4\times10^8$~yr) clusters, because a characteristic disruption timescale of $\tau_*\approx2\times10^8$~yr should be observable as significant flattening or curvature in the mass function of clusters at this age, above the completeness limit of the data. We also use the mass function to assess previous suggestions for a physical upper mass limit of $M_C\approx1$--$2\times10^5~M_{\odot}$ with which clusters in M51 can form (e.g., Gieles et~al.\ 2006b; Gieles 2009). A third goal is to investigate the ages of clusters located in two ``feathers,'' stellar features found between the spiral arms, which are expected, based on simulations, to have ages of $\approx10^8$~yr.

This paper is organized as follows. Section~2 presents the data, cluster selection, and photometry used in this work. Section~3 describes our technique for determining the masses and ages of the clusters, and 
Section~4 presents the mass function for intermediate age (1--$4\times
10^8$~yr clusters) in M51 for two samples, a shallower sample covering
the entire galaxy, and a deeper one covering a $\approx3\times7~\mbox{kpc}^2$ area with low background and little crowding, located on the west side of the galaxy (hereafter referred to as Region~1). Section~5 discusses implications of the shape of the mass function in terms of the formation and disruption of the clusters, and also presents some new results for the ages of clusters in two ``feathers'' located within Region~1. Section~6 summarizes the main results of this work.
 
\section{OBSERVATIONS AND CLUSTER SELECTION}

\begin{figure}
\plotone{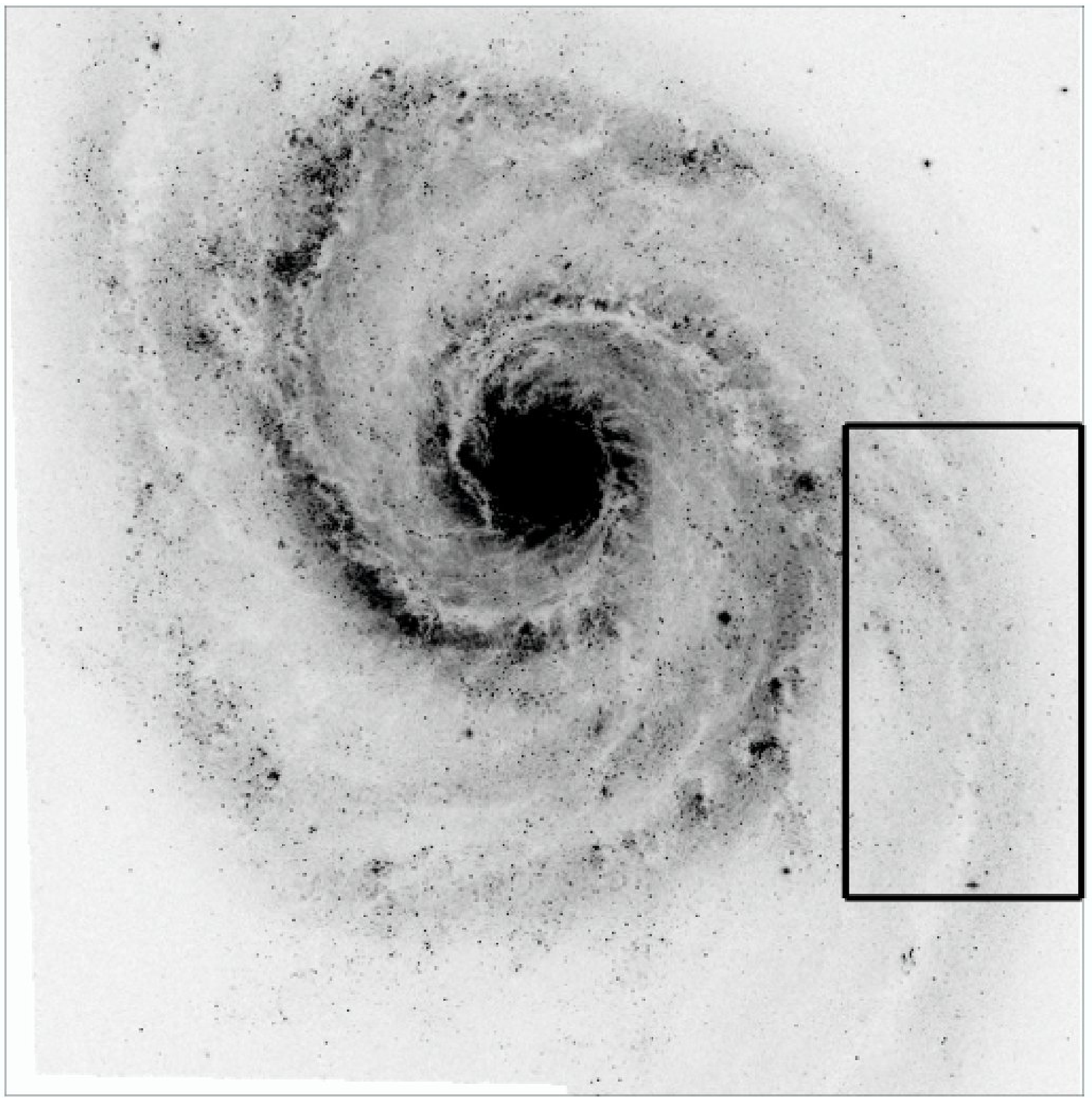}
\caption{Image of M51 with Region~1 marked.}
\label{fig:reg1}
\end{figure}

Observations of M51 (i.e., NGC~5194, a Sbc galaxy) and its companion 
NGC~5195 (a SB0 galaxy) were made in January 2005 with the \textit{Hubble Space Telescope} (\textit{HST}) using the Wide Field Channel of the 
Advanced Camera for Surveys (ACS/WFC), as part of \textit{HST}'s 15th year on-orbit celebration (Program GO-10452; PI: S.~Beckwith). Imaging was obtained in the following optical filters: F435W (``$B$''), F555W (``$V$''), and F814W (``$I$'') and the F658N filter which targets H$\alpha+$[NII] emission at the recessional velocity of M51 (referred to here as the ``H$\alpha$'' filter). Details of the data and image processing are given
at a dedicated webpage: {\tt http://archive.stsci.edu/prepds/m51/}, and result in a mosaic image in each filter with a pixel scale of 0.05\arcsec/pix, or 2~pc/pix at the assumed distance of 8.4~Mpc for M51 (distance modulus $m-M = 29.62$; Feldmeier et~al.\ 1997).  A portion of the resulting image is shown in Figure~\ref{fig:reg1}.

\begin{figure}
\plotone{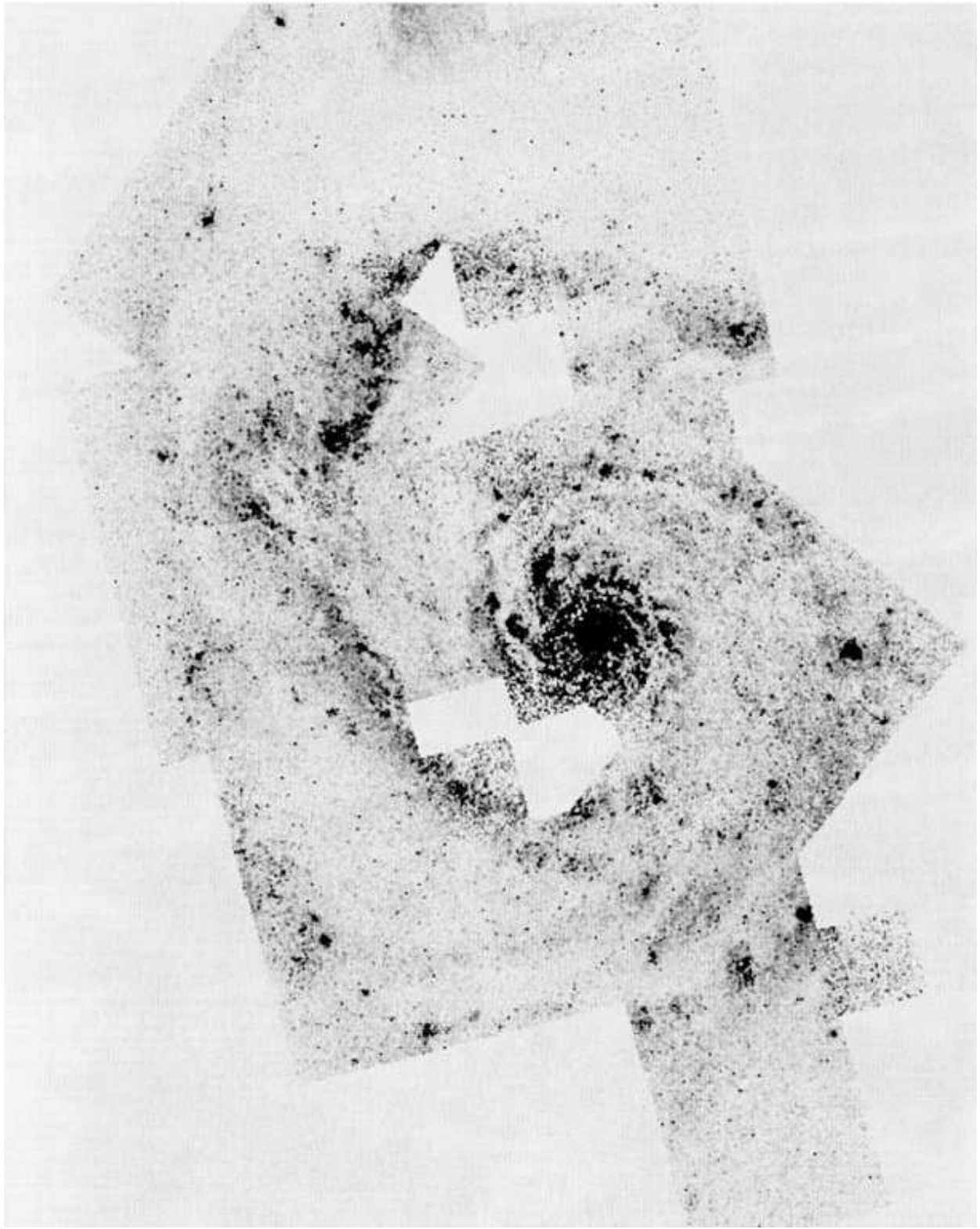}
\caption{$U$-band mosaic of M51 from WFPC2 data.}
\label{fig:uband}
\end{figure}

We have also obtained images for six pointings in M51 with the F336W filter (``$U$'') of the WFPC2 camera as part of program GO-10501 (PI: 
R.~Chandar), which was designed to work in concert with the Hubble Heritage program.  There are two additional pointings available with the F336W filter, which cover the nuclear region of M51 (GO-5652, PI: 
R.~Kirshner and GO-7375, PI: N.~Scoville). The raw data were processed through the standard WFPC2 pipeline (CALWP2) using the most recent
flatfield, bias and dark current reference frames. The resulting calibrated images were co-added to create a single mosaic image using the MULTIDRIZZLE task to reject cosmic rays, identify bad pixels, and correct the
geometric distortion. The final mosaic was created using data primarily
taken with the Wide Field (WF) CCDs,\footnote{We excluded the WF4 chip of the dataset U9GA0601B from the mosaic, because it is badly affected by the WF4 CCD bias anomaly.} with Planetary Camera (PC) data, which have a different scale, used only to cover regions where WF data are absent. Our 
$U$-band mosaic is shown in Figure~\ref{fig:uband}. The resolution of the 
$U$-band mosaic is 0.1\arcsec~pix$^{-1}$ corresponding to 4 pc~pix$^{-1}$, and covers $\approx60$\% of the luminous portion of the ACS mosaic, including some portions of Region~1.

We detect sources, both point-like and slightly extended, on a summed 
\textit{BVI} image,  and perform circular aperture photometry on the drizzled mosaic image in each filter using the IRAF task PHOT, using an aperture radius of 2.5~pixels and a background annulus between 10 and 
13~pixels. For the narrow-band H$\alpha$ image, we perform photometry on the image without subtracting the stellar continuum flux. We convert the instrumental magnitudes to the VEGAMAG magnitude system by applying the zeropoints given in Sirianni et~al.\ (2005) for the ACS filters, and given in Holtzman et~al.\ (1995) for the WFPC2. The F336W photometry was also corrected for the effects of charge transfer inefficiency, using the prescription given by Dolphin (2000; see {\tt http://purcell.as.arizona.edu/wfpc2\_calib/} for the more recent formulae).
The full catalog contains $\approx236,000$ sources, and includes individual
stars, close blends of a few stars, star clusters, and a few background
galaxies. We use two different approaches to determine aperture corrections
for our sources: (1)~we apply a typical value to all clusters; and (2)~we determine an individual aperture correction based on the size measured for the objects, specifically their concentration index ($C$, defined as the 
difference in the aperture magnitudes determined using a 3~pix and a 0.5~pix radius), similar to the procedure used in Chandar et~al.\ (2010b).

In addition to the concentration index, we measure the size of each source
in our catalog using the {\em Ishape} software (Larsen 1999). While the $C$ index is a very simple method that works well for most clusters, it provides a much cruder measure of object size than {\em Ishape}. {\em Ishape} convolves analytic profiles of different effective radii, $R_\mathrm{eff}$, representing the surface brightness profile of a cluster, with the PSF, and 
determines the best fit to each source. It returns a measure of the FWHM of a source in pixels, i.e., how much broader a source is than the PSF. In general, {\em Ishape} can distinguish sources that are broader than the PSF by $\approx0.2$~pixels or more (e.g., Larsen 1999). We created a PSF for the $V$-band image from $\approx30$ relatively bright, isolated point sources, and assume a King profile (King 1966) with a ratio of tidal to core radius of 30. {\em Ishape} returns the best fit radius of each source,
where the fit is performed within a 5~pixel radius.

We construct two different catalogs of star clusters. First, we select clusters over the entire image, but restrict our sample to objects brighter than $V\approx23$~mag, because the contamination due to blends increases significantly below this luminosity in the most crowded spiral arm and high background inner regions of M51. Second, we focus on a large,\break $\approx3\times7~\mbox{kpc}^2$ area of M51 with low background, which also contains many intermediate age (1--$4\times10^8$~yr) clusters (see 
Section~3). In this area, shown by the rectangle in Figure~\ref{fig:reg1}, we are able to select star clusters down to $V\approx24.7$~mag, and hence to construct the mass function down to significantly lower masses.

In order to select star clusters in M51 from our full source catalog, which contains mostly individual stars and clusters, we follow the general approach 
described in Whitmore et~al.\ (2010) and Chandar et~al.\ (2010b), where we construct ``training sets'' of stars and clusters and use their measured properties to guide the criteria used to automatically select clusters. In addition to selecting good clusters,  we need to minimize contamination in
our candidate cluster catalog from close pairs of individual stars. We use the following criteria to select cluster candidates in M51:
(i)~$1.1 \leq C \leq 2.0$; 
(ii)~$0.2 \leq \mbox{FWHM} \leq 2.0$~pix;
(iii)~eliminate sources which satisfy (i) and (ii) but have another detection within 2~pixels; and 
(iv)~select the brightest object within a 5~pixel radius which satisfies (i) and (ii).
These criteria are similar to those used to select clusters in WFC3 observations of M83 (Chandar et~al.\ 2010b). Visual inspection of candidate clusters confirms that we select nearly all obvious clusters down to our magnitude limit, while including relatively few contaminants (mostly blends of a few stars, at the $\approx10$--15\% level) in our candidate cluster list.
The full catalog contains a total of $\approx3500$ cluster candidates brighter than $m_V\approx 23$, and Region~1 contains $\approx2400$ clusters brighter than $m_V\approx 24.7$. The catalogs contain clusters from a few$\times10^6$~yr to $\approx10^{10}$~yr, based on age estimates made in Section~3. We note that our cluster catalog in Region~1 is deliberately biased
{\em against} the youngest clusters, which are preferentially located in and near the inner spiral arms, but instead contains a rich population of intermediate age clusters, making it ideal for the purposes of this study. The youngest clusters ($\tau \lea 10^7$~yr) from our M51 catalog are used in Calzetti et~al.\ (2010) to investigate the upper end of the stellar IMF, and the entire sample will be used to determine the cluster age distribution in a forthcoming study (Chandar et~al.\ 2011, in prep).

\begin{figure}
\plotone{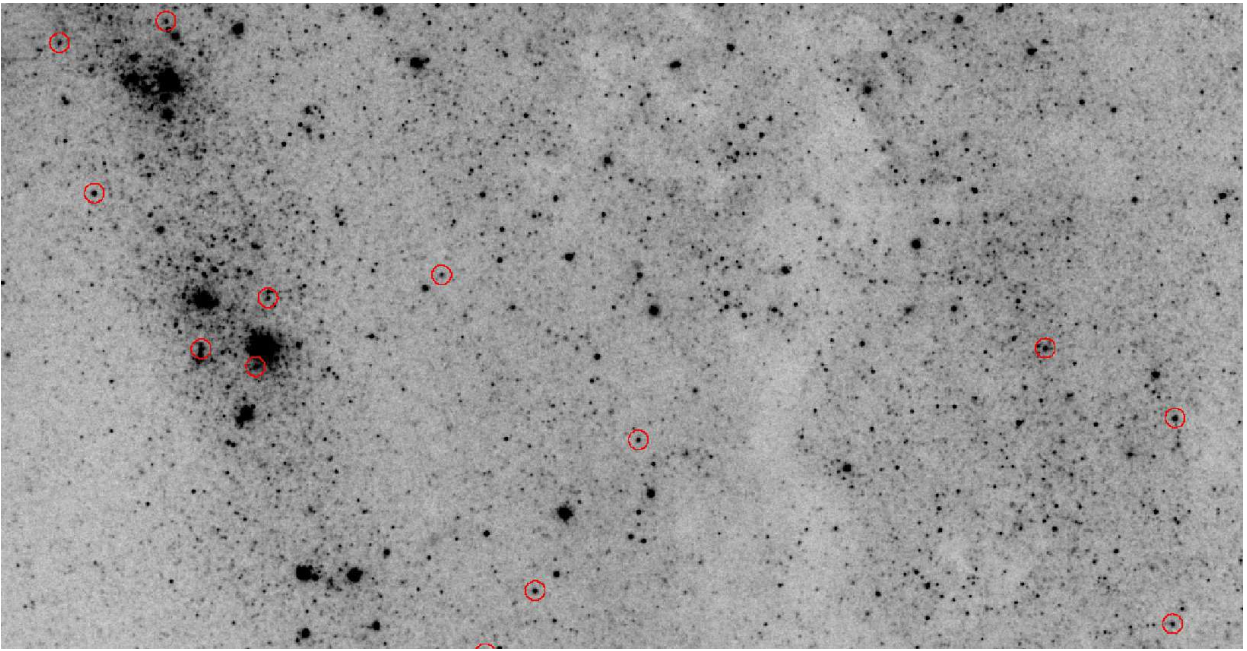}
\caption{
Figure showing a portion of Region~1. Some intermediate age ($\tau\approx1$--$4\times10^8$~yr) clusters with estimated masses of $\approx4$--$8\times10^3~M_{\odot}$ are circled. Clusters with these properties are easy to
detect and clearly broader than the PSF in the ACS images.}
\label{fig:lowmass}
\end{figure}

Incompleteness in our cluster catalog disproportionately affects faint clusters, which are easier to miss in the data. We assess the completeness of our sample by adding artificial clusters throughout the image, and then subjecting them to our selection criteria. These experiments indicate that our sample is fairly complete (at $\approx90$\% level) for a typical cluster size down to $m_V\approx 24.7$ in Region~1,  but that a similar level of completeness occurs closer to $m_V\approx 23$ in the crowded spiral arms.
Our 90\% completeness limit corresponds roughly to a mass of $6\times
10^3~M_{\odot}$ for a cluster at an age of $\tau \approx 2\times10^8$~yr.
Figure~\ref{fig:lowmass} shows that clusters with masses $\approx6$--$8\times10^3~M_{\odot}$ and ages $\approx1$--$3\times10^8$~yr are quite easy to detect in Region~1.

\section{MASS AND AGE DETERMINATIONS}

We estimate the age ($\tau$) and extinction $A_V$ of each cluster as we have done in previous works (see Fall et~al.\ 2005 and Whitmore et~al.\ 2010
for details), by performing a least $\chi^2$ fit comparing the measurements in five filters (\textit{UBVI},H$\alpha$) with predictions from the Bruzual \& Charlot (2007, private communication; see also Bruzual \& Charlot 2003) stellar population models. We have assumed solar metallicity $Z=0.02$ (appropriate for young clusters in M51; Moustakas et~al.\ 2010), a Salpeter (1955) stellar IMF, and a Galactic-type extinction law (Fitzpatrick 1999). The best-fit values of $\tau$ and $A_V$ are those that minimize
\begin{equation}
\chi^2(\tau,A_V) = \sum_{\lambda}
W_{\lambda}~\left(m_{\lambda}^\mathrm{obs}
- m_{\lambda}^\mathrm{mod}\right)^2,
\end{equation}
for each cluster, where $m_{\lambda}^\mathrm{obs}$ and 
$m_{\lambda}^\mathrm{mod}$ are the observed and model magnitudes respectively, and the sum runs over all five bands, $\lambda=U,B,V,I$, and H$\alpha$. In regions where no $U$-band imaging is available, ages are determined from the \textit{BVI},H$\alpha$ filters. The weight factors in the formula for $\chi^2$ are taken to be $W_{\lambda} = 
[\sigma_{\lambda}^2 + (0.05)^2]^{-1}$, where $\sigma_{\lambda}$ is the formal photometric uncertainty determined by PHOT for each band. The mass of each cluster is estimated from the observed $V$-band luminosity, corrected for extinction, and the (present-day) age-dependent mass-to-light ratios ($M/L_V$) predicted by the models.

Our inclusion of photometry in the narrow-band H$\alpha$ filter directly in our fitting procedure (currently unique among studies of extragalactic star cluster systems) helps to break the degeneracy between age and extinction that exists in the broad-band colors of young star clusters.  Most important for this work, the narrow-band filter provides a fourth photometric measurement for clusters where there is no $U$-band imaging available, including portions of Region~1. The narrow-band filter not only provides photometric measurements for clusters that have HII regions and hence strong H$\alpha$ emission, but also for older clusters, since it contains both
(nebular) line and (stellar) continuum emission. In clusters older than 
$\approx10^7$~yr this filter gives a measure of the continuum in the 
$\approx R$ band.

In Chandar et~al.\ (2011, in prep), we provide a detailed assessment of the uncertainties in our age estimates, compare our photometric ages with those derived using other techniques for various cluster subsamples, and
provide basic information on the most massive clusters in M51. Here, we compile a list of clusters with estimated ages between $1$ and $4\times
10^8$~yr, the age range relevant for this work, based on measurements in all five \textit{UBVI},H$\alpha$ filters. We compare these ages with ages derived from different filter combinations. We find that the ages of $\approx85$--90\% of these clusters are recovered to within 0.3~dex when determined from measurements in \textit{BVI}H$\alpha$, $\approx70$--75\% are recovered from measurements in \textit{UBVI}, and only $\approx50$\% are recovered when \textit{BVI} measurements are used with our dating method (in the last case approximately half the clusters are assigned a younger age and relatively high extinction).

Next, we discuss uncertainties in our mass determinations. We have assumed a Salpeter IMF; if we had adopted the more modern Chabrier (2003) IMF instead, the $M/L_V$ and hence the masses would decrease by a near constant (age-independent) $\approx40$\%.\footnote{Figures~2 and
5 in Bruzual \& Charlot (2003) show that there is very little variation in the predicted $M/L_V$ among three different prescriptions for stellar evolution, which include different assumptions regarding convective overshooting and
the detailed treatment of evolved phases of stellar evolution, among others.
The Bruzual \& Charlot models include coded prescriptions for mass lost from the cluster over time due to stellar evolution (assumed to be swept out of a cluster), while compact remnants are assumed to stay within clusters.}
The difference between assuming mean and size-dependent aperture corrections can result in relatively large variations in the total magnitude for a given cluster (for our age determinations we use aperture magnitudes,
which lead to small uncertainties in cluster colors). However, in Section~5 we find that the specific method used to determine aperture corrections
has a negligible effect on the mass function itself.

\section{MASS FUNCTION}

The shape of the mass function for intermediate age star clusters provides one of the most important constraints on the disruption of the clusters.
Figure~\ref{fig:mf} shows the mass function, $dN/dM$, for intermediate age ($8.0 \leq \log(\tau/\mbox{yr}) \leq 8.6$) clusters in Region~1 (top panel), and in the full sample (bottom panel). The mass function reaches lower masses in Region~1 than in the full sample because of the lower background and crowding, as discussed in Section~2. We have used masses determined from cluster luminosities based on our preferred, size-dependent aperture corrections, although the results are quite similar if we use total magnitudes based on an average aperture correction instead (see below).
Two different binnings are shown: variable size bins with equal numbers of clusters in each bin (as recommended by Maiz Apellaniz \& Ubeda 2005 and shown as filled circles), and approximately equal size bins with variable numbers of clusters in each bin (open circles).

These mass functions can be described by a power law, $dN/dM \propto 
M^{\beta}$. We perform simple least square fits to the mass functions, which give formal fits of $\beta=-2.02\pm0.09$ when using variable size bins, nearly identical to the value of $\beta=-2.00\pm0.06$ found from equal size bins for $M \gea 6\times10^3~M_{\odot}$ for intermediate age clusters in Region~1. If we use masses determined when an average aperture correction is applied to the cluster photometry instead of size-dependent aperture corrections, we find nearly identical results,  with $\beta=-2.05\pm0.09$ (variable size bins) and $\beta=-2.03\pm0.06$ (equal size bins). For the full sample and variable size bins, the fits are $\beta=-2.25 \pm 0.09$
and $\beta=-2.24 \pm 0.20$ when using size-dependent and average aperture corrected luminosities and masses, respectively. Hence $\beta$ is nearly the same for the two different samples, within the uncertainties. We find an overall value of $\beta=-2.1\pm0.2$ for the mass function of 1--$4\times10^8$~yr clusters in M51, the mean and approximate range of the fits
determined for Region~1 and for the entire portion of M51 covered by the ACS observations.

Our results for the mass function of star clusters in M51 are in good agreement with the earlier one from Bik et~al.\ (2003), who found a power-law index of $\beta=-2.1\pm0.3$ for clusters in M51 based on shallower WFPC2 observations, and within the uncertainties of the values of 
$\beta=-1.68\pm0.33$ (for clusters with $8.0 \leq \log(\tau/\mbox{yr}) < 8.3$) and $\beta=-1.81\pm0.27$ (for clusters with $8.3 \leq \log(\tau/\mbox{yr}) < 8.5$) found by Hwang \& Lee (2010). It is different however, from the steep value of $\beta=-2.76\pm0.28$ (for clusters more massive than $6\times 10^4~M_{\odot}$ and with ages of $\tau=1$--$6\times10^8$~yr)
found by Gieles (2009), but similar to $\beta=-2.09\pm0.09$ for $10^7$--$10^8$~yr clusters with $M \gea 10^4~M_{\odot}$ found by Gieles (2009).

\section{DISCUSSION}

Here we use the observed shape of the mass function of 1--$4\times10^8$~yr clusters in M51 to investigate two issues related to the formation and disruption of the clusters. First, we assess whether or not our observations support previous suggestions for a physical upper mass limit with which
clusters in M51 can form. Next, we assess the validity of the specific 
mass-dependent disruption model that has been claimed for clusters in M51,
and discuss the implications of our results for the galaxy-dependent disruption picture for star clusters. Finally, we investigate the ages of clusters in two ``feathers'' located within Region~1.

\subsection{Constraints on an Upper Mass Cutoff}

The upper end of the mass functions of old globular star clusters are typically better fit by a Schechter function, $\psi(M) \propto M^{\beta}\mbox{exp}^{-M/M_C}$, with a cutoff of $M_C\approx1$--$2\times10^6~M_{\odot}$, than by a power law (e.g., Burkert et~al.\ 2000; Fall \& Zhang 2001;
Jordan et~al.\ 2007). Some recent works have suggested that a Schechter function may also better describe the mass function of young $\tau \lea \mbox{few} \times 10^8$~yr clusters in nearby spiral galaxies, but with a lower cutoff of  $M_C\approx 2\times10^5~M_{\odot}$ (e.g., see Portegies Zwart et~al.\ 2010 and references therein).

\begin{figure}
\epsscale{0.8}
\plotone{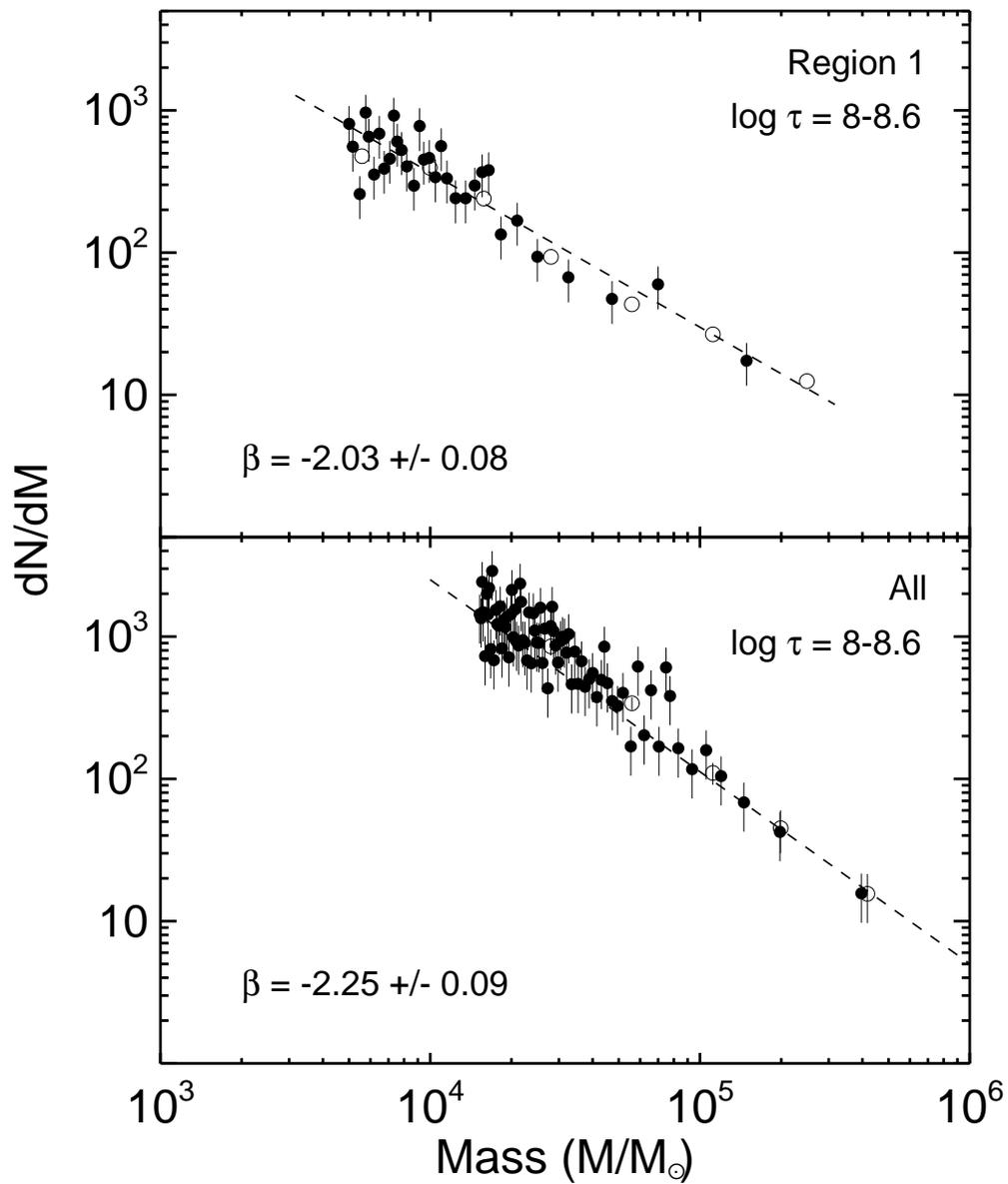}
\caption{Mass function for log~$\tau=8.0$--8.6~yr clusters in M51 in Region~1 (top panel) and in the full area covered by the ACS images (bottom panel).
The solid circles show the results for variable bin widths and the open circles show fixed bin widths. The best fit power-law index $\beta$ for variable-size
bins is given. See text for details.
}
\label{fig:mf}
\end{figure}

Figure~\ref{fig:mf} compared the observed mass function of clusters in M51 with a power law, and the upper panel of Figure~\ref{fig:mfdatamodels} compares the observed mass function with Schechter functions for three different values of $M_C$: $2\times10^5~M_{\odot}$, $4\times10^5~M_{\odot}$, and $10^6~M_{\odot}$. These figures suggest that a single power-law provides a better fit to the observations than a Schechter function with $M_C=2\times10^5~M_{\odot}$, because there are too many massive star clusters present compared with predictions from the latter.
More quantitatively, when we compare a power-law with the best fit index from Figure~\ref{fig:mf} of $\beta=-2.03$ ($\beta=-2.25$) with the mass function for clusters with $M\geq 10^5~M_{\odot}$, a K-S test returns a 
$P$-value of 0.74 (0.92) for Region~1 (all of M51). Hence a power law provides an acceptable fit (has a $P$-value $> 0.05$) to the upper end
of the mass function, with no significant deviation either visually or statistically. Our data therefore, do not prefer a Schechter function over a power law.

If we do fit a Schechter function to the observed mass function, we find a best fit of $\beta=-2.01$ and $M_C \approx10^{12}~M_{\odot}$, i.e., essentially a power law. If we assume a Schechter function with $\beta=-2.0$, we can rule out $M_C$ values of $2\times10^5~M_{\odot}$ with high statistical confidence, since a K-S test returns a $P$-value of only $4\times 10^{-4}$, with lower values of $M_C$ giving even lower $P$-values. Any value of $M_C > 2\times10^5~M_{\odot}$ therefore gives a statistically acceptable fit to the observations, within a 95\% confidence interval. This is consistent with the $M_C\approx1$--$2\times10^6~M_{\odot}$ found for old globular clusters.

This is a different conclusion than that reached by Gieles et~al.\ (2006b) and Gieles (2009), who suggested that the mass function for clusters in M51 is better fit by a Schechter function with $M_C\approx2\times10^5~M_{\odot}$ than by a power law.\footnote{We make similar assumptions here as the Gieles et~al.\ works when estimating cluster masses and ages, namely a Salpeter IMF and a Galactic-type extinction law.} However, the Gieles et~al.\ results were based on WFPC2 observations that have limited coverage of M51.
We find that these WFPC2 observations miss nearly $2/3$ of the 50 most massive clusters with ages between 1--$4\times 10^8$~yr found in our full sample, including the few most massive objects. This may indicate that small number statistics, and hence the absence of a few massive clusters,
can significantly affect tests for a Schechter-like cutoff. We found a similar result for clusters in M83, another grand-design spiral galaxy (Chandar et~al.\ 2010b), where a power law provides a good fit to the observed mass function, without requiring an exponential cutoff (Chandar et~al.\ 2010b).
Taken together, our results do not support previous suggestions for a universal
cutoff $M_C$ around $\approx1$--$2\times 10^5~M_{\odot}$ for spiral galaxies.

\subsection{Constraints on Mass-Dependent Disruption}

Some processes, such as the evaporation of stars due to internal relaxation,
are known to disrupt lower mass clusters earlier than higher mass clusters.
Mass-dependent disruption processes can be detected directly, because they lead to curvature or flattening in the mass function of clusters at lower masses.

\begin{figure}
\epsscale{0.8}
\plotone{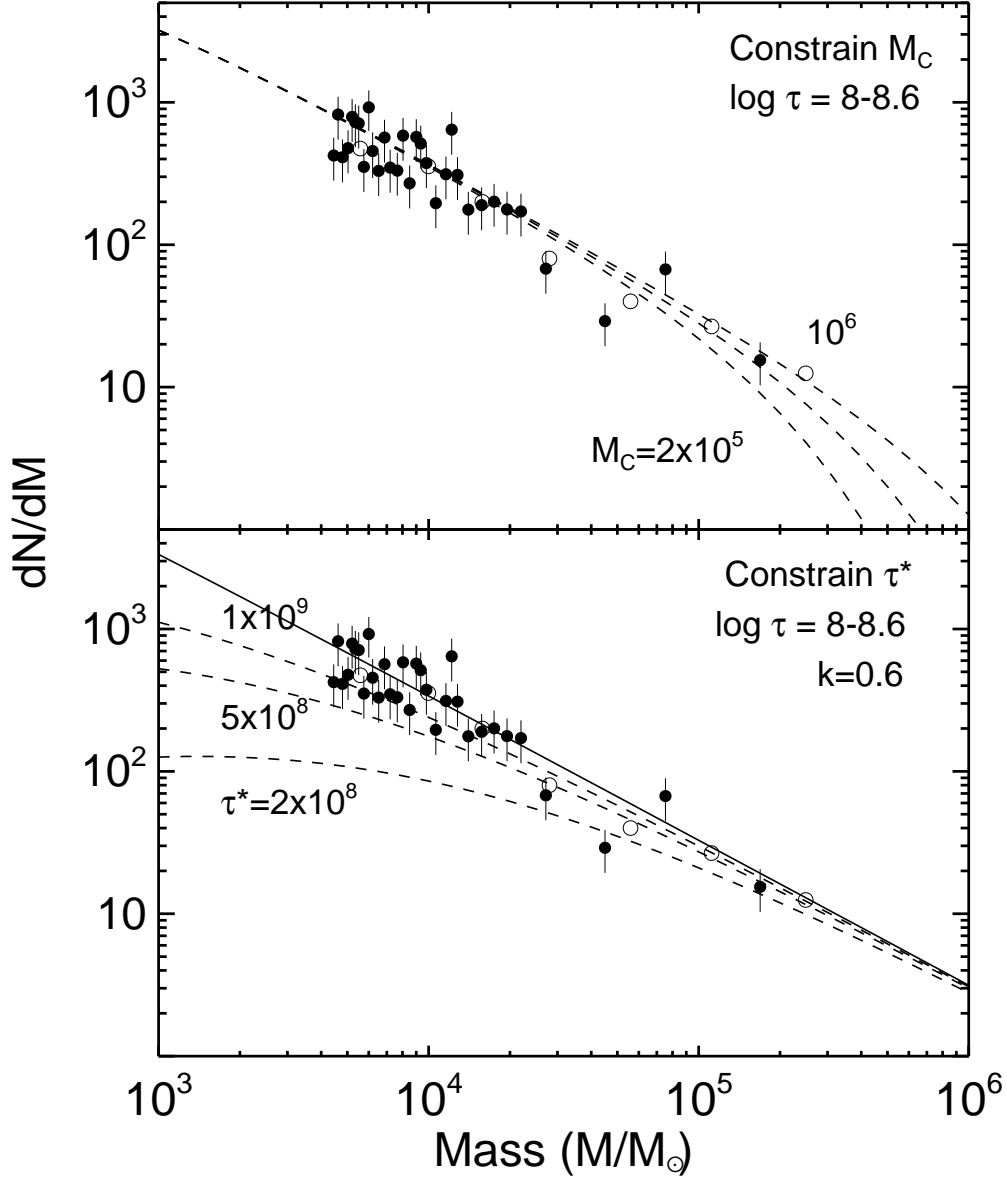}
\caption{Mass function for 1--$4\times10^8$~yr clusters in Region~1 is compared with predictions for an upper mass cutoff $M_C$ (upper panel) and for different characteristic disruption timescales $\tau_*$ (lower panel).
The solid circles show the results for variable bin widths and the open circles show fixed bin widths. The curves in the upper panel are Schechter functions,
with $\psi(M) \propto M^{\beta} \mbox{exp}(-M/M_C)$, for $M_C=2\times10^5~M_{\odot}$, $4\times10^5~M_{\odot}$, and $1\times10^6~M_{\odot}$ (dashed curves). The curves in the lower panel use equations~(B6)--(B8) from Fall et~al.\ 2009 to predict the evolution of the mass function for a population of log~$\tau=8.0$--8.6~yr clusters, with the indicated disruption
timescales $\tau_{*}$, for an assumed exponent $k=0.6$.  See text for details.}
\label{fig:mfdatamodels}
\end{figure}

The mass function presented for Region~1 in Figure~\ref{fig:mf} is the deepest to date for intermediate age clusters in M51, and hence well-suited
for establishing the presence or absence of bends. In the bottom panel of Figure~\ref{fig:mfdatamodels} we compare the observed mass function for intermediate-age clusters in M51 with predictions from the gradual, mass-dependent cluster disruption model suggested by Boutloukos \& Lamers (2003), with formulae derived by Fall et~al.\ (2009). The model assumes that the mass $M$ of a cluster evolves with time $\tau$ according to the equations
\begin{equation}
dM/d\tau = - M/\tau_d(M),
\end{equation}
\begin{equation}
\tau_d(M) = \tau_* (M/M_*)^k,
\end{equation}
where the exponent $k$ and characteristic disruption timescale $\tau_*$ are adjustable parameters, while $M_*=10^4~M_{\odot}$ is a fiducial mass scale.
We assume that the initial clster mass function is a power law with $\beta=-2.0$.\footnote{This assumption is consistent with the results found by Hwang \& Lee (2010) for the mass function of $\tau \lea 10^7$~yr clusters in M51.} Equations~(B6)--(B8) in Fall et~al.\ (2009) give the analytic expressions derived for the mass function of star clusters in different intervals of age in the case of gradual, mass-dependent disruption, and can be compared directly with observations. The lowest curve, which shows 
$\tau_*=2\times10^8$~yr and $k=0.6$, The values suggested by Gieles 
et~al.\ (2006a), predicts significantly more curvature than is observed in the data.

Previous works reached a different conclusion regarding the disruption of clusters in M51 than we found here. We believe this is due to the fact that these works applied the indirect method developed by Boutloukos \& Lamers (2003), which averages clusters over large ranges of mass and age and assumes that any bends observed in the mass and age distributions are due to mass-dependent disruption. However, bends can appear for other reasons, such as artifacts related to systematic errors or statistical noise, and
are expected to appear in any case when the mass function is unrestricted
in age (see Fall et~al.\ 2009 for details).

When no bends or curvature are present in the mass function, only a lower limit can be determined  for $\tau_*$. We use the observed mass function and the predicted curves shown in the bottom panel of Figure~\ref{fig:mfdatamodels} to estimate a lower limit for $\tau_*$ of $\approx1\times10^9$~yr, for an assumed value of $k=0.6$. Higher values of $k$ would lead to more curvature than shown for $k=0.6$, and we find a lower limit of 
$\approx 2\times10^9$~yr for an assumed value of $k=1$ (not shown). In summary, our results do not support previous suggestions of a characteristic disruption timescale $\tau \approx 1$--$2 \times10^8$~yr for a typical
$10^4~M_{\odot}$ cluster in M51.

\subsection{Implications for the Galaxy-Dependent Cluster Disruption Picture}

We have just shown that the power-law shape observed for the mass function only allows us to place a lower-limit on $\tau_*$, i.e., that the mass-dependent disruption of clusters in M51 is not observed directly over the mass-age range probed here. This result is similar to those found previously for clusters in several different galaxies, including the Antennae, the LMC and SMC, M83 and M51. These galaxies span a factor of 100 in star formation rate, have different total masses and morphologies, and include spirals, dwarf irregulars and merging galaxies. Together, they represent a good sample to assess whether or not there are strong variations in the rate of dynamical evolution of star clusters in different galaxies, at least for the first $\approx$few$\times10^8$~yr. We have not found direct evidence in any of these systems that lower mass clusters are disrupted faster than higher mass clusters, although our observations are sensitive to different ranges in mass and age in each galaxy. Based on these results, we conclude that current observations do not support the notion that the disruption timescale $\tau_*$ varies strongly for young clusters from galaxy to galaxy,
and hence our results do not support the galaxy-dependent disruption picture.
In a forthcoming paper, we will study the age distribution of clusters in M51 at different masses, and the mass function at different ages, to assess whether or not these support the ``quasi-universal'' picture for cluster disruption.

\subsection{Ages of Clusters in Feathers}

\begin{figure}
\epsscale{0.7}
\plotone{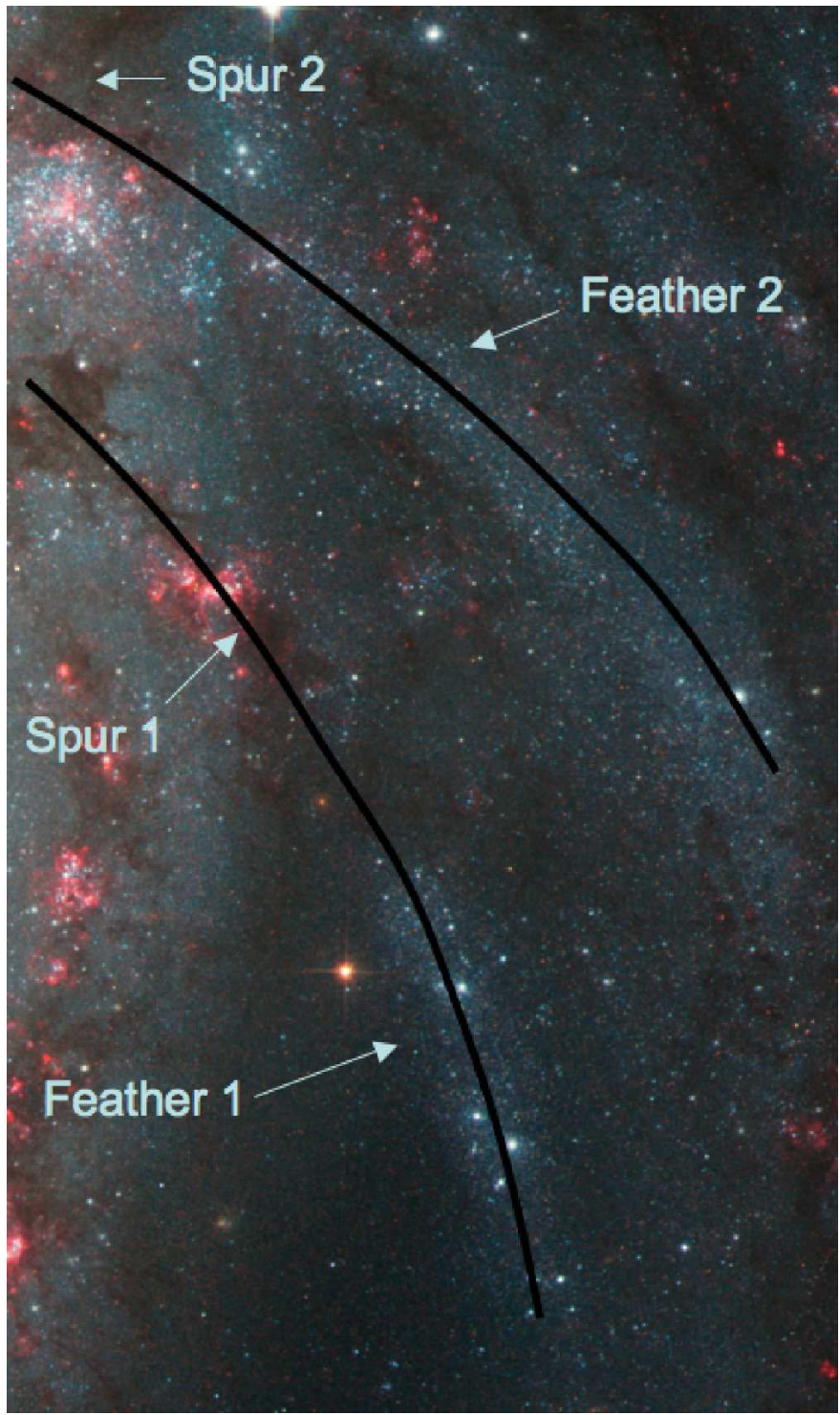}
\caption{Location of the two spurs and feathers in M51 discussed in this work.}
\label{fig:featherlocation}
\end{figure}

There are two interesting stellar features located within Region~1, and shown in Figure~\ref{fig:featherlocation}.  These features, which we refer to as ``feathers,'' contain a number of star clusters, have larger pitch angles than the spiral arm and appear to be a continuation of dark extinction features (we refer to these extinction features as ``spurs''). The spurs are also detected in CO (e.g., Corder et~al.\ 2008; Koda et~al.\ 2009), and appear to emanate from the outer (leading) portion of the spiral arm, suggesting that they are related in some way to both the spiral arm and to the feathers. The spurs are also labeled in Figure~\ref{fig:featherlocation}.

\begin{figure}
\epsscale{1}
\plotone{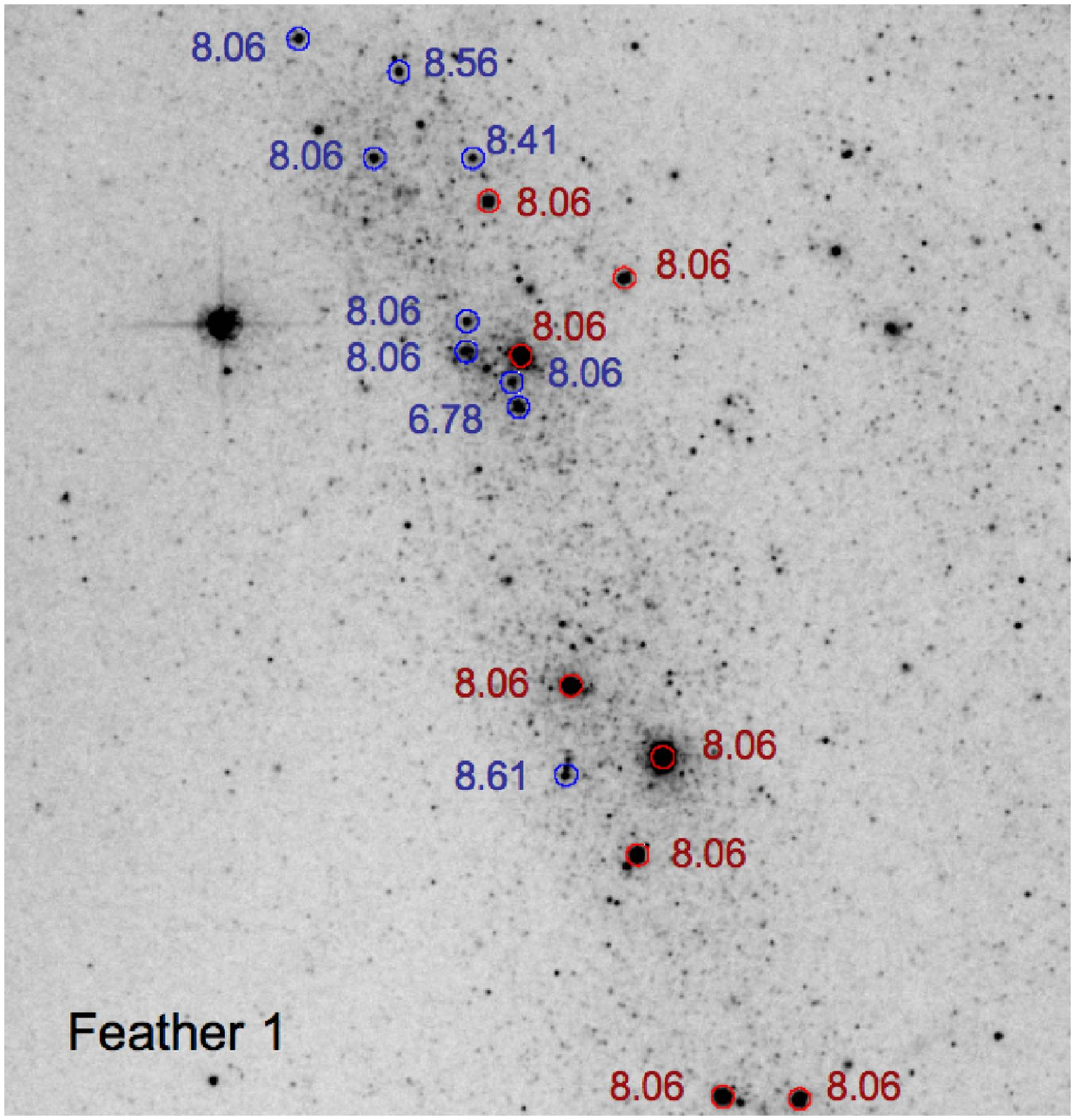}
\caption{
a) Image of Feather~1, located within Region~1, with age estimates of clusters brighter than $M_V\approx-7$ (red) and $M_V\approx-6$ (blue) labeled. }
\label{fig:feather1}
\end{figure}

Figure~\ref{fig:feather1}a shows that clusters brighter than $m_V\lea 22$ ($M_V\lea -7$; shown in red) located in feather~1 virtually all have the {\em same} age, and formed $\approx 1\times10^8$~yr ago. Most of the fainter clusters, with $m_V$ between $-6$ and $-7$ (shown in blue) have similar estimated ages, although fainter clusters have larger uncertainties in their ages. We also note that the $\approx$25th most massive, intermediate age star cluster in M51 is located within feather~1.

Figure~\ref{fig:feather2}b shows the locations and estimated ages for clusters brighter than $M_V\lea -7$ (shown in red) in feather~2. These clusters have similar but unidentical ages, with most forming $\tau \approx2\times10^8$~yr ago. Fainter clusters, shown in blue, show a similar trend in their ages as the brighter clusters, although there are a handful of fainter clusters with significantly younger age estimates, where we believe our dating technique erroneously found a younger age and higher extinction than for the other clusters. When compared with their counterparts in feather~1, most clusters in feather~2 appear to be somewhat {\em older}, by a factor of $\approx2$, and also to have formed over a longer period.
The formal uncertainties in our age determinations for clusters in the feathers are only log~$(\tau/ \mbox{yrs})\approx0.05$--0.1, supporting the idea of different ages for clusters in the two feathers. The age difference comes from a difference in the observed colors of the clusters. Over a similar magnitude range, clusters in feather~2 have a redder median \textit{V-I} color than clusters in feather~1 by $\approx0.15$~mag, and also have a larger spread in \textit{V-I} color ($\sigma=0.12$ vs.\ $\sigma=0.05$ for feather~1). We do not see evidence for age gradients along either feather, as might be expected if star formation had been triggered at one end and propagated across to the other end. In addition to forming earlier, 
feather~2 also has a larger pitch angle than feather~1.

The locations of $\approx10^8$~Myr old clusters found here can help to distinguish between different scenarios for the excitation of spiral structure
in M51. Based on a number of previous simulations, Dobbs \& Pringle (2010) predict the locations of clusters with different ages from four mechanisms which can generate spiral structure:
(i)~a global, steady density wave;
(ii)~a rotating, central bar;
(iii)~local gravitational instabilities; and
(iv)~strong, external tidal interaction.
In these simulations, Dobbs \& Pringle (2010) track the trajectory of the dense gas, and assume that these trajectories are not dissimilar to those of stars. Notional clusters with ages $\approx10^8$~Myr are found primarily between spiral arms, including in feather-like structures with relatively large pitch angles, in the simulations for a global density wave and in a tidally induced spiral, but {\em not} for flocculent spirals. A central bar has not been observed in M51, implying that this model is also unlikely to explain spiral structure in M51, although the simulations do show intermediate age clusters in feather-like features with appropriate pitch angles. We will make a stronger test of how spiral structure is induced in M51 in a future paper, by comparing the locations of clusters of all ages younger than $\approx \mbox{few} \times 10^8$~yr across M51 with results from the Dobbs \& Pringle (2010) simulations.

\section{SUMMARY}

We have used \textit{UBVI},H$\alpha$ images taken with the ACS and WFPC2 cameras on-board the \textit{Hubble Space Telescope} to study star clusters in the nearby grand-design spiral galaxy M51. We estimated the masses and ages of the clusters by comparing the measured photometry with predictions from the population synthesis models of Bruzual \& Charlot (2007).

We constructed the mass function of intermediate age (1--$4\times10^8$~yr)
star clusters in a $3\times 7$~kpc$^{2}$ region of M51, and found that it is 
well described by a single power law, $dN/dM \propto M^{\beta}$, with 
$\beta=-2.02\pm 09$, for $M \gea 6\times10^3~M_{\odot}$. This mass function is deeper by a factor of $\approx5$ than those available in the literature. We also found $\beta=-2.25\pm0.09$ for the mass function of intermediate age clusters across all of M51, but only down to masses of 
$\approx10^4~M_{\odot}$ due to the difficulty of selecting clusters in the crowded inner and spiral arm regions. This gives an average value of $\beta=-2.1\pm0.2$. The mass function does not have obvious bends at either 
the low or high mass end. We found that a power-law provides a statistically acceptable fit to the mass function, based on quantitative tests, and does not require an upper mass cutoff. Therefore, we do not confirm previous suggestions that there is a cutoff of $M_C\approx2\times10^5~M_{\odot}$ for
clusters in M51.  

We also assessed previous suggestions for mass-dependent disruption of clusters in M51, where clusters are disrupted on timescales that depend on their mass as $\tau_d(M) = \tau_* (M/M_*)^k$, with the specific values $k=0.6$ and $\tau_*=2\times10^8$~yr for a fiducial $M_*=10^4~M_{\odot}$ cluster. These specific values predict a significant amount of curvature which is not observed in our mass function. We concluded that previous works reached an incorrect conclusion because they applied an indirect method which makes the {\em a~priori} assumption that clusters are disrupted in a manner that depends on their initial mass, to lower quality observations with poorer spatial coverage taken with \textit{HST}/WFPC2.
The deeper observations presented here, that cover M51 nearly in its entirety, did not show evidence that lower mass clusters are disrupted earlier than higher mass cluster in M51, at least over the studied range of masses and ages.

Our results for the mass function of star clusters in M51 are similar to those found recently for $\tau \lea \mbox{few} \times 10^8$~yr cluster systems in several other galaxies, including the Magellanic Clouds, M83, and the Antennae, where we also did not find evidence for mass-dependent disruption. In light of these results, we concluded that there is currently no clear observational support for the suggestion that young clusters in different galaxies disrupt on very different timescales and in a mass-dependent fashion.

We also studied the ages of clusters in two prominent ``feathers'' located in Region~1, stellar structures beyond the inner spiral arm, which have a larger pitch angle than the arms. The clusters located within one feather appear to have formed coevally, approximately $1\times10^8$~yr ago, while clusters in the second feather formed earlier, with a typical age of $2\times10^8$~yr, but over a longer period. We compared the locations of 
$\approx10^8$~yr clusters in M51 with predictions from Dobbs \& Pringle 2010, and ruled out a scenario where self-gravity of the stellar disk is responsible for generating the spiral structure in M51.

\acknowledgments
R.~C.\ is grateful for support from NSF through CAREER award 0847467 and from NASA through grant GO-10501-01-A from STScI, which is operated by AURA, Inc., under NASA contract NAS5-26555.

\clearpage

\clearpage

\clearpage

\clearpage

\clearpage

\clearpage

\clearpage

\clearpage

\addtocounter{figure}{-1}
\begin{figure}
\plotone{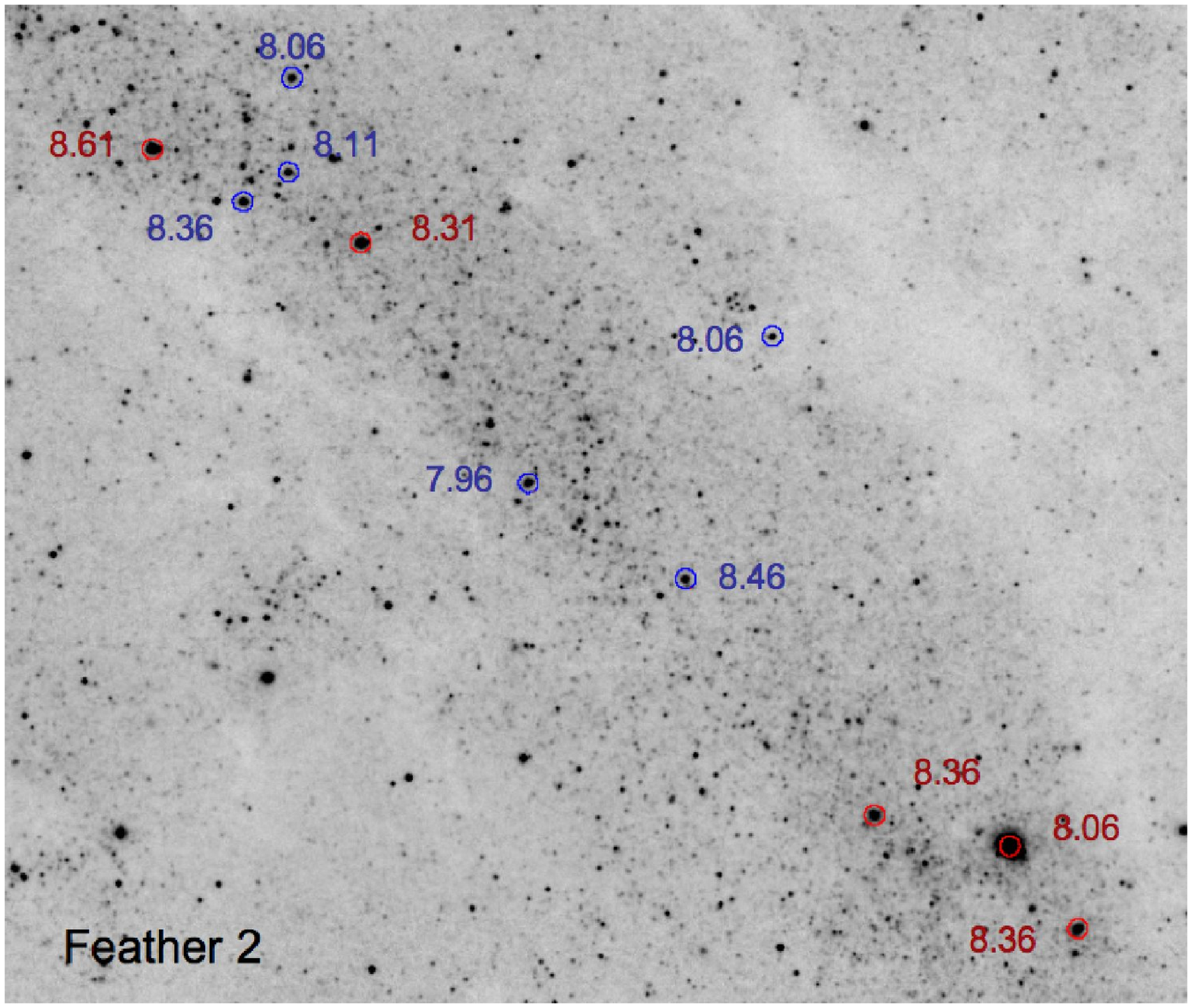}
\caption{b) Same as Figure~7a, but for Feather~2.}
\label{fig:feather2}
\end{figure}

\end{document}